\documentclass[12pt]{article}
\begin{document}

\title{Path integral measure in Regge calculus from the
 functional Fourier transform}
\author{V.M.Khatsymovsky \\
 {\em Budker Institute of Nuclear Physics} \\ {\em
 Novosibirsk,
 630090,
 Russia}
\\ {\em E-mail address: khatsym@inp.nsk.su}}
\date{}
\maketitle
\begin{abstract}
The problem of fixing measure in the path integral for the
Regge-discretised gravity is considered from the viewpoint
of it's "best approximation" to the already known formal
continuum general relativity (GR) measure. A rigorous
formulation may consist in comparing functional Fourier
transforms of the measures, i.e. characteristic or
generating functionals, and requiring these to coincide on
some dense set in the functional space. The possibility for
such set to exist is due to the Regge manifold being a
particular case of general Riemannian one (Regge calculus
is a minisuperspace theory). The two versions of the
measure are obtained depending on what metric tensor,
covariant or contravariant one, is taken as fundamental
field variable. The closed expressions for the measure are
obtained in the two simple cases of Regge manifold. These
turn out to be quite reasonable one of them indicating that
appropriately defined continuum limit of the Regge measure
would reproduce the original continuum GR measure.
\end{abstract}
\newpage
Regge calculus still remains the most natural discrete
regularisation of general relativity (GR) promising from
the viewpoint of constructing well-defined quantum gravity
theory \cite{RegWil}. The result of quantisation being
expressed in the form of the path integral, the key
question is that of the choice of the integration measure.
In particular, the earliest quantum formulation of 3D Regge
calculus \cite{PonReg} is based on specific property of
$6j$-symbols whose product for large values of arguments
reduces to a kind of the path integral with the Regge
calculus action, the arguments of $6j$-symbols being
interpreted as linklengths. There are a number of models
generalising these results to the physical 4D case
\cite{RegWil}, of which the Barrett-Crane one \cite{BarCra}
attracts much attention for it analogously reproduces path
integral with the Regge calculus action \cite{BarWil}.

The path integral measure in numerical simulations is
usually chosen as the simplest among the invariant ones
\cite{HamWil}. Normalising the measure w.r.t. the DeWitt
supermetric would allow to fix the measure uniquely
\cite{MenPei}. However, in the 4D case this construction
turns out to suffer from unrenormalisable UV divergences
provided by singular nature of Regge manifold;
discretisation of the Faddeev-Popov ghost field improves
the situation, but the measure turns out to be singular at
the point of superspace  of metrics corresponding to the
flat spacetime \cite{Kha1}. The reason for this is rather
simple and connected with the change of gauge content of
the theory in the flat spacetime when a certain variations
of linklengths become gauge ones since they do not change
geometry. The same unpleasant feature displays also in the
canonical quantisation of the (3+1)D (continuous time)
Regge calculus \cite{Kha2}. The singularity of the measure
at the flat spacetime makes extracting physical
consequences from the theory a difficult task because of
the absence of the perturbative expansion around the flat
spacetime. Therefore it may be useful to study the problem
of quantum Regge calculus within another framework. Thus
far Regge calculus has been treated as independent theory
without any reference to the continuum GR. Now consider it
as simply approximation to or regularisation of the already
quantised continuum GR. So we need to define a notion of
"the best approximation" to the known formal expression for
the continuum measure
\begin{equation}
\label{d-mu-C}                                         
d\mu_{\rm C} = \prod_x{({\rm det}\|g_{ik}\|)^{-{5\over 2}}
d^{10}g_{ik}}.
\end{equation}
(This simplest local invariant measure can be shown to
correspond to the canonical quantisation of GR in a certain
gauge,
\cite{KonPop}\footnote{Also the papers had been appeared,
refs. \cite{FraVil,Kak,AraChe} where the arguments in
favour of the measure $d\mu_{\rm C}\!$ = $\!\prod_x{
({\rm det}\|g_{ik}\|)^{-3/2}g^{00}d^{10}g_{ik}}$ instead of
the eq.(\ref{d-mu-C}) used here were given. It was shown in
that papers that the difference between the both
expressions amounts to the new purely renormalisation terms
in the perturbation theory for gravity without changing the
structure of the theory.}.)
This notion can be given a strict sense by treating the
measure as a functional $\int (\cdot )d\mu$ on the space of
the functionals of metric. Since Regge metric is a
particular case of general Riemannian one, the functional
on Riemannian metrics can be viewed at the same time as
that on Regge ones. Thus, the two measures, continuum
(\ref{d-mu-C}) and discrete Regge one of interest
$d\mu_{\rm R}$
can be defined on the same set of metric functionals.
Looking for such set dense in the space of metric
functionals in the appropriate topology and requiring that
both measures would coincide on this set we can define
$d\mu_{\rm R}$. The exponents of linear functionals of
metric
("functional plane waves") just present such set, probably
the only one which gives possibility to have expressions
definable and calculable on the functional level.

The above approach is natural also from
the axiomatic point of view where a functional subject to
a certain set of axioms, the Osterwalder-Shrader ones,
can be considered as functional Fourier transform of a
measure of some quantum field theory \cite{GliJaf}, the
so-called {\it characteristic functional} usually referred
to in the physical literature as {\it generating
functional}. The analog of the characteristic functional
considered in our case takes the form
\begin{equation}
\label{mu-C-hat}                                       
\hat{\mu}_{\rm C}(f) = \int {e^{ig(f)}d\mu_{\rm C}(g)}
\end{equation}
where there are the two possibilities for the linear metric
functional $g(f)$,
\begin{equation}
\label{g1}                                             
\int {f_{ik}g^{ik}\sqrt {g}d^4x} \equiv g_1(f)
\end{equation}
or
\begin{equation}
\label{g2}                                             
\int {f^{ik}g_{ik}\sqrt {g}d^4x} \equiv g_2(f)
\end{equation}
depending on what metric tensor, $g^{ik}$ or $g_{ik}$, is
chosen as true field variable; the $f_{ik}(x)$ or
$f^{ik}(x)$ is probe function (since quantum fields are
generally treated as distributions, the probe functions are
usually supposed to be infinitely differentiable with
compact supports). Strictly speaking, the measure involved
in the definition of the characteristic functional should
include also $\exp {(-S)}$, the $S$ being the (Euclidean)
gravity action. Occurence of this factor would make the
explicit calculations not easier than defining and
calculating the gravity path integral itself. Therefore we
are trying to define Regge analog $d\mu_{\rm R}$ of the
$d\mu_{\rm C}$
separately from $\exp {(-S)}$. A point of view on omitting
this factor within strict framework of characteristic
functional may consist in saying that the strong coupling
limit ($S\!$ $\rightarrow\!$ 0) is considered.

Thus, our approach to definition of the Regge measure
$d\mu_{\rm R}$ amounts to setting
\begin{equation}
\label{mu-C-mu-R}                                      
\hat{\mu}_{\rm R}(f_{\rm R}) = \hat{\mu}_{\rm C}(f_{\rm R})
\end{equation}
on a discretised version $f_{\rm R}$ of the probe
functions. The
only natural choice for the tensor $f_{\rm R}$ on Regge
manifold
is to take it being piecewise-constant in the
piecewise-affine frame, that is, constant on each the
4-simplex whenever $g_{ik}$ is constant on it. Then one
tries to define $\hat {\mu}_{\rm C}(f_{\rm R})$ (where
$f_{\rm R}$ is not
smooth but is a limit of smooth functions).

Strictly speaking, the measure $d\mu_{\rm C}$ does not
exist as
mathematical object, a regularisation is implicit. There
should be some care with this regularisation. For example,
the measure $d\mu_{\rm C}$ looks formally positive, and
regularisation should keep this property. Convenience of
the characteristic functional is, in particular, just that
the positivity property looks rather simple if written in
terms of this functional,
\begin{equation}
\label{positive}                                       
\sum^N_{\alpha,\beta=1}{c_{\alpha}\bar{c}_{\beta}
\hat{\mu}_{\rm C}(f_{\alpha}-\bar{f}_{\beta})}\geq 0
\end{equation}
for any sequence of the probe functions $f_\alpha$ and
complex numbers $c_\alpha$, $\alpha\!$ = $\!1$, $\!2\!$,
..., $\!N$. If then $d\mu_{\rm R}$ is defined via
(\ref{mu-C-mu-R}), it's positivity immediately follows from
(\ref{positive}). So we imply that positivity of the
measure, if required, is ensured in the continuum GR; then
it is guaranteed for our construction of the Regge measure
too.

Now turn to our characteristic functional (\ref{mu-C-hat})
which proves to be the product over points,
\begin{equation}
\label{muC-PiIx}                                       
\hat{\mu}_{\rm C}(f) = \prod_x{I(x)},
\end{equation}
of the factors (for $g(f)\!$ = $\!g_1(f)$)
\begin{equation}
\label{int1}                                           
I = I_1 = \int{e^{if_{ik}g^{ik}\sqrt{g}d^4x}({\rm det}
\|g_{ik}\|)^{-{5\over 2}+\epsilon}d^{10}g_{ik}}.
\end{equation}
Here a nonzero $\epsilon$ is introduced because, as
mentioned above, the measure $d\mu_{\rm C}$ does not exist
without regularisation, therefore not specifying the latter
this measure can be understood whenever this is possible in
the sense of analytical continuation from the sufficiently
large positive $\epsilon$ where (\ref{int1}) can be defined
to the point $\epsilon\!$ = $\!0$ of interest; the $d^4x$
is an infinitesimal "bare" (i.e. corresponding to the
Euclidean metric $g_{ik}\!$ = $\!\delta_{ik}$) 4-volume
associated to a point. To calculate this (and $I\!$ =
$\!I_2$ for $g(f)\!$ = $\!g_2(f)$) note that
\begin{eqnarray}
({\rm det}\|g_{ik}\|)^{-{5\over 2}+\epsilon}d^{10}g_{ik}
& = & ({\rm det}\|g^{ik}\sqrt{g}\|)^{-{5\over 2}+\epsilon}
d^{10}(g^{ik}\sqrt{g})\nonumber\\
& = & ({\rm det}\|g_{ik}\sqrt{g}\|)^{-{5\over 2}+{\epsilon
\over 3}}d^{10}(g_{ik}\sqrt{g}).                       
\end{eqnarray}
Perform the following change of variables,
\begin{equation}
g^{ik}\sqrt{g} = \sum_A{e^i_A\lambda_Ae^k_A}          
\end{equation}
where $e^i_A\!$ = $\!0$ at $A\!$ $>$ $\!i$, $e^i_A\!$ =
$\!1$ at $A\!$ = $\!i$ (this is the Gaussian decomposition
of the symmetrical matrix into the product of a diagonal
and twice a triangular one with unity diagonal elements).
Integral turns out to be Gaussian over $d^6e^i_A$ and
factorisable over $d^4\lambda_A$. Physical region of
positivity of $g^{ik}\sqrt{g}$ is picked out by
inequalities $\lambda_A\!$ $>$ $\!0$ $\forall$ $\!A$.
Correspondingly, over $\lambda_A$ the cosine (sine) Fourier
transform in that region is adequate,
\begin{eqnarray}
I_1 & \Rightarrow & \prod^4_{A=1}{\left\{\int^{\infty}_0{
\lambda_A^{{3\over 2}+\epsilon -A}d\lambda_A\int^{+\infty}
_{-\infty}\cdots\int^{+\infty}_{-\infty}{2\cos{[\lambda_A
(e^i_Af_{ik}e^k_A)d^4x]}\prod_{i>A}{de^i_A}}}\right\}}
\nonumber\\
& = & (4\pi)^4(2d^4x)^{-4\epsilon}\Gamma(2\epsilon-1)
\Gamma(2\epsilon-3)\cos^2{\!\pi\epsilon\,}({\rm det}f)
^{-\epsilon}                                          
\end{eqnarray}
(accidentally, the same follows for sine transform).
Important is only proportionality to $({\rm det}f)
^{-\epsilon}$ and that inversed of the same operation
reduced to Regge lattice should be applied to recover Regge
measure of interest. Expression for $I_2$ corresponding to
the choice $g(f)\!$ = $\!g_2(f)$ follows by replacing
$\epsilon$ by $\epsilon/3$ and substituting $f$ by
$f^{ik}$.

Now consider reduction of $\exp{(ig(f))}$ and
$\hat{\mu}_{\rm C}(f)$ to the Regge lattice; the functional
relating these two reduced objects will be just the
discrete measure of interest. The piecewise-affine frame is
fixed by attributing the coordinates $x^i_a$, $i\!$ =
$\!1$, $\!2$, $\!3$, $\!4$ to each vertex $a$. The length
squared $s_{(ab)}$ of the link $(ab)$ connecting the
vertices $a$ and $b$,
\begin{equation}
\label{s-g}                                           
s_{(ab)} = l^i_{ab}l^k_{ab}g_{ik}(\sigma),~~~l^i_{ab}
\equiv x^i_a - x^i_b,
\end{equation}
is a particular example of the so-called edge components
$f_{(ab)}$ of a symmetrical second rank tensor $f_{ik}$
constant in a simplex $\sigma$ \cite{PirWil},
\begin{equation}
\label{f-edge-cov}                                    
f_{(ab)} = l^i_{ab}l^k_{ab}f_{ik}(\sigma).
\end{equation}
Here $g_{ik}(\sigma)$ and $f_{ik}(\sigma)$ are the values
of the tensors in a simplex $\sigma$ containing the link
$(ab)$. The edge components unambiguously parameterise a
symmetrical rank two covariant tensor in a 4-simplex. But
whereas
$s_{(ab)}$ does not depend on the choice of $\sigma$
containing the link $(ab)$, the $f_{(ab)}$ may do so.
However, the number of variables $f_{(ab)}$
Fourier-conjugate to the metric should be the same as the
number of independent variables $s_{(ab)}$ parameterising
the metric, i. e. the number of links. Therefore the
condition is required that the variables $f_{(ab)}$ should
not depend on $\sigma$ $\!\supset$ $\!(ab)$. The
possibility to
have $f_{(ab)}$ constrained by this condition becomes
evident if one imagines that $f_{(ab)}$ are the new squared
linklengths of our Regge manifold instead of $s_{(ab)}$,
the scheme of linking and coordinates of vertices being the
same (some of these linklengths can be made imaginary, if
necessary, in the sense of analytical continuation); the
metric tensor
in the piecewise-affine frame in the 4-simplices of thus
constructed Regge lattice will be just $f_{ik}(\sigma)$.
Then according to the ref. \cite{PirWil} the functional
(\ref{g1}) on Regge lattice functions takes the form
\begin{equation}                                      
g_1(f_{\rm R}) = 2\sum_{\sigma}{\sum_{(ab)\subset\sigma}{
f_{(ab)}{\partial V_{\sigma}\over\partial s_{(ab)}}}}
= 2\sum_{(ab)}{f_{(ab)}{\partial V\over\partial s_{(ab)}}}.
\end{equation}
Here $V_{(\sigma)}$ is the volume of $\sigma$, $V\!$ = $\!
\sum\nolimits_{\sigma}{V_{\sigma}}$ is the volume of the
manifold (in the compact case).

Analogously, let $f^{(ab)}$ be independent variables
living on the links. Let us define the contravariant
symmetrical rank two tensor $f^{ik}$ constant inside each
the 4-simplex $\sigma$,
\begin{equation}
\label{f-edge-con}                                    
f^{ik}(\sigma) = \sum_{(ab)\subset\sigma}{f^{(ab)}l^i_{ab}
l^k_{ab}}.
\end{equation}
Using this anzats we get
\begin{equation}                                      
g_2(f_{\rm R}) = \sum_{(ab)}{f^{(ab)}s_{(ab)}V_{(ab)}}.
\end{equation}
Here we have introduced notation for a volume associated to
a link,
\begin{equation}
V_{(ab)} = \sum_{\sigma\supset (ab)}{V_{\sigma}}.     
\end{equation}

Next reduce the expression (\ref{muC-PiIx}), (\ref{int1})
to the Regge lattice when $f(x)$ is piecewise-constant,
$f(x)\!$ = $\!f(\sigma)$ whenever $x\!$ $\in$ $\!\sigma$.
Then $I(x)$ is piecewise-constant too, and for $g(f)\!$ =
$\!g_1(f)$ we have
\begin{equation}
\label{I-on-Regge}                                    
\prod_x{I(x)} = \prod_{\sigma}{\prod_{x\in\sigma}
{I_1(\sigma)}} = \prod_{\sigma}{I_1(\sigma)^{N_{\sigma}}}
\sim \prod_{\sigma}{({\rm det}f(\sigma))^{-\epsilon
N_{\sigma}}}
\end{equation}
(and analogously for $g(f)\!$ = $\!g_2(f)$ with the
replacement $\epsilon\!$ $\rightarrow$ $\!\epsilon/3$)
where only dependence on $f$ is shown. Here $N_{\sigma}$
is a number of points contained in a simplex $\sigma$; of
course, the continuum measure is defined in the limit
$N_{\sigma}\!$ $\rightarrow$ $\!\infty$ starting from the
originally finite $N_{\sigma}$. If integration over metric
is made, information on the simplex size is lost, and the
only choice symmetrical w.r.t. the different simplices and
points is to consider $N_{\sigma}$ being equal to the same
value $N$ for all the $\sigma$'s. Then, if we keep $N$
finite before taking the limit $\epsilon\!$ $\rightarrow\!$
0, we can redefine $N\epsilon\!$ $\rightarrow$ $\!\epsilon$
(or $N\epsilon/3\!$ $\rightarrow$ $\!\epsilon$ for $g(f)$ =
$g_2(f)$), so that
\begin{equation}
\label{mu-hat-Regge}                                  
\hat{\mu}_{\rm C}(f_{\rm R}) \sim \prod_{\sigma}{({\rm det}
f(\sigma))^{-\epsilon}}.
\end{equation}
This corresponds to the naive idea that the product over
points should turn, up to a normalisation factor, into the
same product but over simplices. Also we observe that it is
namely the measure (\ref{d-mu-C}) for which this
correspondence takes place; were the exponent there
different from $-5/2$, as in the footnote following the eq.
(\ref{d-mu-C}), the reduction to the Regge lattice
like (\ref{I-on-Regge}) could not be defined in such the
simple way. Take into account parameterisation of the
tensors $f_{ik}$, $f^{ik}$ in terms of the edge components
(\ref{f-edge-cov}), (\ref{f-edge-con}). Then
\begin{equation}                                      
{\rm det}\|f_{ik}(\sigma)\| = ({\rm det}\|l^i_{ab}\|)^{-2}
\Delta_1(f; \sigma)
\end{equation}
where $\Delta_1(f; \sigma)$ is the so-called bordered
determinant \cite{Sor} composed of the variables $f_{(ab)}$
living on the links $(ab)$ belonging to $\sigma$. Note that
$(\Delta_1(s; \sigma))^{1/2}\!$ = $\!V_{\sigma}$, the
volume of $\sigma$. The $\|l^i_{ab}\|$ means the matrix of
any four link vectors $l^i_{ab}$ of the simplex not
laying in the same 3-plane. In the case $f\!$ = $\!f^{ik}$
we can find even more simple expression,
\begin{equation}                                      
{\rm det}\|f^{ik}(\sigma)\| = ({\rm det}\|l^i_{ab}\|)^2
\Delta_2(f; \sigma),~~~\Delta_2 \equiv \sum^{~~~~~,}
_{(a_ib_i)\subset\sigma}{f^{(a_1b_1)}f^{(a_2b_2)}
f^{(a_3b_3)}f^{(a_4b_4)}}
\end{equation}
where the summation runs over all the unordered
combinations of the four links of the simplex, $(a_ib_i)$,
$i\!$ = $\!1$, $\!2$, $\!3$, $\!4$ not laying in the same
3-plane.

Finally, we write out up to the normalisation factor the
relation which fixes the Regge measure if we choose for the
fundamental metric field the $g^{ik}$,
\begin{equation}
\label{mu1}                                           
\int{\exp{\left(2i\sum_{(ab)}{f_{(ab)}{\partial V\over
\partial s_{(ab)}}}\right)d\mu^{(1)}_{\rm R}(s)}} =
\prod_{\sigma}{(\Delta_1(f; \sigma))^{-\epsilon}},
\end{equation}
or $g_{ik}$,
\begin{equation}
\label{mu2}                                           
\int{\exp{\left(i\sum_{(ab)}{f^{(ab)}s_{(ab)}V_{(ab)}}
\right)d\mu^{(2)}_{\rm R}(s)}} = \prod_{\sigma}{(\Delta_2
(f; \sigma))^{-\epsilon}}.
\end{equation}

Consider the simplest case of Regge manifold consisting of
the two identical 4-simplices $\sigma_1$, $\sigma_2$ with
mutually identified vertices. Then $f_{ik}(\sigma_1)\!$ =
$\!f_{ik}(\sigma_2)$ or $f^{ik}(\sigma_1)\!$ =
$\!f^{ik}(\sigma_2)$ if parameterised by $f_{(ab)}$ or
$f^{(ab)}$ according to (\ref{f-edge-cov}) or
(\ref{f-edge-con}). Now using $f_{(ab)}$ (or $f^{(ab)}$)
and $f_{ik}$ (or $f^{ik}$) as Fourier conjugate variables
is equally convenient, because the number of links $(ab)$
coincides with the number 10 of the components of $f_{ik}$
(or $f^{ik}$) taken in one of the simplices. So we do not
need to parameterise tensors by the edge components and can
write immediately
\begin{equation}
\label{mu-g-ik-con}                                   
\int{\exp{\left(i{2\over 4!}f_{ik}g^{ik}\sqrt{g}{\rm det}
\|l^i_{ab}\|\right)}d\mu^{(1)}_{\rm R}(g)} = ({\rm det}
\|f_{ik}\|)^{-2\epsilon}
\end{equation}
or
\begin{equation}
\label{mu-g-ik-cov}                                   
\int{\exp{\left(i{2\over 4!}f^{ik}g_{ik}\sqrt{g}{\rm det}
\|l^i_{ab}\|\right)}d\mu^{(2)}_{\rm R}(g)} = ({\rm det}
\|f^{ik}\|)^{-2\epsilon}
\end{equation}
instead of (\ref{mu1}), (\ref{mu2}) (again, up to a
normalisation factor). Here we have taken into account that
the product in the RHS consists of the two identical
factors, so the exponent is simply rescaled, $\epsilon\!$
$\rightarrow$ $\!2\epsilon$. The inverse Fourier transform
is then straightforward and gives
\begin{equation}
\label{d-mu-g-ik}                                     
d\mu^{(1)}_{\rm R} = d\mu^{(2)}_{\rm R} = ({\rm det}
\|g_{ik}\|)^{-5/2}d^{10}g_{ik}
\end{equation}
up to normalisation, or, in terms of linklengths,
\begin{equation}
\label{d-mu-s}                                        
d\mu_{\rm R} = V^{-5}d^{10}s,
\end{equation}
the $V$ being the volume of the simplex, $V\!$ =
$\!(\Delta_1(s))^{1/2}$.

The above example deals with the strongly curved spacetime;
next consider the simplest Regge minisuperspace model of
the flat spacetime. Take the flat 4-parallelepiped with all
it's diagonals emitted from one of it's vertices and
compactified toroidally by imposing periodic boundary
conditions (on the linklengths). This is the simplest,
consisting of 24 4-simplices elementary cell of the
periodic Regge lattice \cite{RocWil} specified here by the
conditions of compactness and flatness. The flatness means
that the linklengths of the body and hyperbody diagonals
can be expressed in terms of the linklengths of the 4
parallelepiped edges and 6 face diagonals. Equivalently,
the metric $g_{ik}$ can be taken the same in all the 24
4-simplices. Since the number of components $g_{ik}$
coincides with the number 10 of independent linklengths, as
in the example above, we again may work not passing to the
variables $s$. Further, if we study the measure on the
Regge minisuperspace constrained by additional conditions
on the linklengths or metric, we need the same number of
the conditions also on the Fourier conjugate variables $f$.
In our case metric $g_{ik}$ being the same in all the 24
equivalent 4-simplices, the Fourier transform of the
measure of interest depends on $f^{ik}(\sigma)$ through the
sum $f^{ik}(\sigma_1)\!$ + $\!f^{ik}(\sigma_2)\!$ +
$\!...\!$ + $\!f^{ik}(\sigma_{24})$. The most symmetrical
way of setting the conditions on $f^{ik}(\sigma)$ is to
equate these for all the 4-simplices, $f^{ik}(\sigma)\!$
$\equiv$ $\!f^{ik}$, and analogously for $f_{ik}$. Finally,
in the RHS we have the product of the 24 identical factors,
so the exponent $\epsilon$ is rescaled to $24\epsilon$,
\begin{eqnarray}                                      
\int{\exp{(if_{ik}g^{ik}\sqrt{g}{\rm det}\|l^i_{ab}\|)}
d\mu^{(1)}_{\rm R}(g)} & = & ({\rm det}\|f_{ik}\|)
^{-24\epsilon},\\
\int{\exp{(if^{ik}g_{ik}\sqrt{g}{\rm det}\|l^i_{ab}\|)}
d\mu^{(2)}_{\rm R}(g)} & = & ({\rm det}\|f^{ik}\|)
^{-24\epsilon}.                                       
\end{eqnarray}
The answer is notationally the same as in the above
example, eq. (\ref{d-mu-g-ik}).

Thus, in the two examples, those of strongly curved and
flat Regge manifolds we have obtained the same expressions
for the measure written in terms of metric. This means that
the measure cannot crucially depend on the curvature. On
the other hand, the example of the flat spacetime might be
relevant to the continuum limit of the Regge calculus.
Indeed, if one triangulates a fixed smooth manifold with
the help of Regge manifolds and tends the maximal
linklength $a$ of these manifolds to zero making
triangulation finer and finer, then the angle defects of
these Regge manifolds tend to zero too as $Ra^2$, $R$ being
typical curvature of the smooth manifold. The result we
have obtained is just the expression for the continuum
measure, although for a specific case when the product over
points runs over only one point.

Despite that the two versions of the Regge measure coincide
in the above simple cases, these are generally different,
and thus far there is no indication which one of them is
preferable.
In general case we can write out explicit expression for
the measure as convolution of elementary measures like
(\ref{d-mu-s}) for all the 4-simplices,
\begin{eqnarray}
\label{mu1R}                                          
d\mu^{(1)}_{\rm R} & = & \prod_{(ab)}{\left[d\left({
\partial V\over\partial s_{(ab)}}\right)\right]}\int{
\left\{\prod_{\sigma}{\Delta_2(h_{(\sigma)}; \sigma)^{-5/2
+\epsilon}}\right.}\nonumber\\
& & \cdot\left.\prod_{(ab)}
{\left[\delta\left(\sum_{\sigma\supset (ab)}{h
_{(\sigma)}^{(ab)}} - {\partial V\over s_{(ab)}}\right)
\prod_{\sigma\supset (ab)}{dh_{(\sigma)}^{(ab)}}\right]}
\right\}
\end{eqnarray}
or
\begin{eqnarray}
\label{mu2R}                                          
d\mu^{(2)}_{\rm R} & = & \prod_{(ab)}{\left[d(s_{(ab)}
V_{(ab)})\right]}\int{\left\{\prod_{\sigma}{\Delta_1
(\tilde{s}^{(\sigma)}; \sigma)^{-5/2+\epsilon}}\right.}
\nonumber\\
& & \cdot\left.\prod_{(ab)}
{\left[\delta\left(\sum_{\sigma\supset (ab)}{\tilde{s}
^{(\sigma)}_{(ab)}} - s_{(ab)}V_{(ab)}\right)\prod_{\sigma
\supset (ab)}{d\tilde{s}^{(\sigma)}_{(ab)}}\right]}\right\}
\end{eqnarray}
where $\tilde{s}^{(\sigma)}_{(ab)}$ and
$h^{(ab)}_{(\sigma)}$ are dummy variables living on the
pairs 4-simplex --- edge. It is taken into account that
$\Delta_1^{-\epsilon}$ and $\Delta_2^{-5/2+\epsilon}$ or
$\Delta_1^{-5/2+\epsilon}$ and $\Delta_2^{-\epsilon}$ are
mutually connected by Fourier transform, as it follows from
the relation of $\Delta_1$ and $\Delta_2$ to determinants
of the co- and contravariant metric.

Probably the crucial
difference between the two versions would display in the 2D
model. There an analog of the eq. (\ref{mu1R}) could not be
derived directly by Fourier transform of the continuum
measure because the functional plane waves as functionals
of $g^{ik}\sqrt{{\rm det}\|g_{ik}\|}$ do not depend on the
conformal degree of freedom of the metric and thus do not
form a dense set. But even being derived via analytic
continuation from the dimensionality $n\!$ $\neq$ $\!2$,
the $d\mu^{(1)}_{\rm R}$ given by eq. (\ref{mu1R}) (where
now $V$ is the total square) is degenerate for it does not
depend on
the differential of the global conformal degree of freedom,
while the $d\mu^{(2)}_{\rm R}$ does so. This can serve as
some argument in favour of the version
$d\mu^{(2)}_{\rm R}$, although the absence of the 2D puzzle
can not be the criterium for the 4D case.

The expressions (\ref{mu1R}), (\ref{mu2R})
remind those for Feynman diagrams in the usual quantum
field theory. However, the role of propagators is played by
the fourth order polynomials raised to the negative
half-integer power and with nontrivial position of zeroes.
This makes analytic evaluation of nontrivial such graph
quite difficult. But prior to that the problem of
regularising the
original continuous measure should be considered. In the
simple examples of the present paper  the explicit form of
this regularisation turns out to be unimportant for the
formal expression for the resulting Regge measure in the
limit $\epsilon\!$ $\rightarrow\!$ 0. In the general case
it is unclear whether expressions (\ref{mu1R}),
(\ref{mu2R}) remain finite at $\epsilon\!$ $\rightarrow\!$
0 without any additional regularising $d\mu_{\rm C}$ or
not; if not, this would mean that the final formal
expressions for $d\mu_{\rm R}$ would generally depend on
this regularisation.

\bigskip
This work was supported in part by the grants E00-33-148
and RFBR No. 00-15-96811.

\end{document}